\documentclass[twocolumn,floatfix, showpacs]{revtex4}
\usepackage{color}

\usepackage[final]{epsfig}
\usepackage{amsmath}
\usepackage{parskip}
\usepackage{wasysym}

\begin{document}

\title{Dimuon transverse momentum spectra as a tool to characterize the emission region in heavy-ion collisions}

\author{Thorsten Renk}
\email{trenk@phys.jyu.fi}
\affiliation{Department of Physics, PO Box 35 FIN-40014 University of Jyv\"askyl\"a, Finland}
\affiliation{Helsinki Institute of Physics, PO Box 64 FIN-00014, University of Helsinki, Finland}

\author{J\"org Ruppert}
\email{ruppert@physics.mcgill.ca}
\affiliation{Department of Physics, 3600 rue University, McGill University, Montreal, Quebec, Canda, H3A 2T8 }


\begin{abstract}
Previous dilepton measurements in heavy-ion collisions have mainly focused on invariant mass spectra to clarify in-medium changes of vector meson properties. However, a dimuon is characterized by two scales --- invariant mass $M$ and transverse momentum $p_T$. Like transverse momentum spectra of hadrons, $p_T$ spectra of dileptons arise from an interplay between emission temperature and collective transverse flow, whereas the invariant mass is insensitive to flow. Having two control parameters of which only one is sensitive to flow allows at given $M$ to characterize the emission region in terms of average temperature and flow. Thus, one is able to study what phases of the fireball evolution radiate into a given mass window. We demonstrate this technique using the dimuon transverse momentum spectra measured by the NA60 collaboration and present strong arguments that a thermalized evolution phase with $T > 170$ MeV leaves an imprint in the spectra.

\end{abstract}

\maketitle

\section{Introduction}
\label{sec_introduction}

The in-medium properties of vector mesons have long been considered a tool to observe chiral symmetry restoration experimentally. In order to provide the medium, decays of vector mesons produced in heavy-ion collisions have to be studied, and there is a long history of model calculations and measurements \cite{RappSummary}. So far, the experimentally accessible quantity has been the invariant mass $M$ spectrum of dileptons, a quantity in which the in-medium spectral function, averaged over the evolution history of the medium, is reflected.

With the NA60 high precision dimuon measurement \cite{NA60data} it has become clear that the averaged in-medium spectral function in the vector channel shows broadening of the $\rho$ meson and no significant mass shift \cite{BRruppert}. 
While it is difficult to establish this broadening as an unambiguous signal for chiral symmetry restoration, a suggestion has been made in \cite{RappNA60} that dilepton emission in the invariant mass region above 0.9 GeV might be dominated by $4\pi$ processes if a considerable chiral mixing of vector and axial vector correlators is assumed.
On the other hand, in \cite{NA60calc} we found that this region is dominated by partonic radiation. Thus, a tool to further characterize the emission region dominating a particular region in the invariant mass spectrum is clearly important before more definite conclusions can be drawn.

Recently, NA60 has presented measurements of transverse momentum $p_T$ spectra in three different mass bins \cite{NA60ptspec}. This allows to study dimuon emission as a function of two scales $M$ and $p_T$ with different properties. While $M$ is invariant under Lorentz transformations and the invariant mass spectrum is hence (apart from detector acceptance effects) insensitive to collective flow of the medium, $p_T$-spectra are heavily influenced by flow.

The importance of collective flow for the description of hadronic $p_T$ spectra in heavy-ion collisions has been realized a long time ago (cf. e.g. \cite{HeinzThermal}). For experimental results from the Relativistic Heavy Ion Collider (RHIC) in Brookhaven, the analysis of both single hadron spectra and two-particle correlations in terms of thermal spectra subject to a flow field has been extremely fruitful to characterize the final state of the evolution \cite{BlastWave}. Extension of the method to a parametrization of the full evolution instead of the final state has been demonstrated to give a good characterization of spectra and two-particle correlations at both SPS 158 AGeV fixed target Pb-Pb collisions \cite{Synopsis} and RHIC 200 AGeV Au-Au collisions \cite{RHIC}. In the following, we use the same techniques, adapted to the different nature of electromagnetic emission processes, to characterize the emission region dominating the $p_T$-spectra in the different mass bins measured by NA60. Conceptually, this is similar to the ideas outlined in \cite{HeinzDiflow} where elliptic flow is instead suggested to characterize the emission region.

\section{Dilepton emission and flow} 

Our model for dilepton emission in InIn collisions is outlined in \cite{NA60calc}, here we reproduce the essential details as far as needed to understand the flow analysis.

We calculate the dilepton spectrum from the convolution 

\begin{eqnarray}
\label{E-1}
\frac{d^3N}{dM dp_T d\eta} = \text{evolution} \otimes \frac{dN}{d^4 x d^4q} 
\end{eqnarray}

where the rate is found from the averaged virtual photon spectral function $R(q, T, \rho_B)$ as  

\begin{eqnarray}
\label{E-Rate}
\frac{dN}{d^4 x d^4q}  =  \frac{\alpha^2}{12\pi^4} \frac{R(q, T, \rho_B)}{e^{p_\mu u^\mu/T}- 1}.
\end{eqnarray}

Here, $q$ represent the four-momentum of the emitted muon pair,
$T$ is the temperature of the emitting volume element and $\rho_B$ its baryon density.
The finite lepton mass $m_L$ can be accounted for by an additional factor of $(1+2 m_L^2/M^2) \sqrt{1-4m_L^2/M^2}$ in Eq. (\ref{E-Rate}). The fireball evolution encodes information on the evolution of radiating volume, temperature $T$ and baryon chemical potential $\mu_B$, local transverse flow velocity $v_T$,  longitudinal rapidity $\eta$ and in the late evolution stage chemical non-equilibrium properties such as pion chemical potential $\mu_\pi$ and Kaon chemical potential $\mu_K$.

The model considers matter with a transverse entropy density $s$ distribution
\begin{equation}
s(r, \tau) = 1/\left(1 + \exp\left[\frac{r - R_c(\tau)}{d_{\text{ws}}}\right]\right)
\end{equation}
where $R_c(\tau) = R_0 + \frac{a_\perp}{2} \tau^2$. Onto this distribution,  a transverse flow field 
\begin{equation}
\label{E-Flow}
\rho_T = \text{atanh } v_\perp(\tau) =  r/R_c(\tau) \cdot \rho_c(\tau)
\end{equation}
is imposed with $\rho_c(\tau) = \text{atanh } a_\perp \tau$. $a_\perp \tau_f = v_f$ is the final velocity at radius $R_C(\tau_f)$ at freeze-out. Note that this is not the maximal velocity reached at the system's surface: The surface is dynamically calculated from the local entropy density and the equation of state via the condition $T=T_F$. At $\tau_f$, a thermalized system exists only close to $R=0$ and $R_c(\tau_f)$ is outside the thermalized region. Since the position of the Cooper-Frye surface changes in a non-trivial way the model cannot completely be characterized by one number for the flow.
We have also investigated flow fields with dependences $\rho \sim r^2$ and $\rho \sim \sqrt{r}$ which are suitably normalized such as to lead to hadronic $p_T$ spectra with the same slope as the form Eq.~(\ref{E-Flow}) (see \cite{Synopsis,RHIC} for details about the calculation of hadronic emission from the model).

Let us recall the role of transverse flow: The factor $p_\mu u^\mu$ in Eq.~(\ref{E-Rate}) evaluates for a process at midrapidity to 
\begin{equation}
\label{E-TraFo}
E = m_T \cosh(\rho_T) - p_T \sinh(\rho_T) \cos(\phi)
\end{equation}
where $\phi$ is the angle between flow direction and dilepton momentum and $E$ the energy of a virtual photon measured with $M, p_T, m_T = \sqrt{M^2 + p_T^2}$ as it appears in the local restframe of a fluid element with radial rapidity $\rho_T$. For a particular choice of flow field and transverse density distribution and a Boltzmann distribution instead of the rate $\frac{R(q, T, \rho_B)}{e^{\beta p_\mu u^\mu}- 1}$ in Eq.~(\ref{E-Rate})  (as appropriate for hadrons), the spatial integral of the convolution of the flow field with density yields analytical expressions for the spectrum \cite{HeinzThermal, BlastWave}, however we solve the integral numerically.

It can be seen as a general property of Eq.~(\ref{E-TraFo}) that the resulting effect on the spectral slope will in general be $p_T$ dependent. This is apparent from the two limiting cases $p_T \ll M$ where the equation becomes $E\approx M\cosh{\rho}$ and $p_T \gg M$ where one finds $E \approx p_T \exp[\rho]$ for $\phi = \pi$ and $E\approx p_T \exp[-\rho]$ for $\phi = 0$. It is well known that the effect of flow makes the spectral slope harder, in the case of large masses and flow values even leading to a flattening of the spectrum, the 'knee' seen e.g. for protons \cite{BlastWave}. This hardening of the slope is mass dependent as long as $p_T$ is of the order of $M$.
 
The slope of a spectum in the model thus arises as follows: The temperature $T$ in Eq.~(\ref{E-Rate}) sets the basic scale possibly modified by the momentum (and density)   dependence of the spectral function $R(q,T,\rho_B)$. We point out that the modifications of the $p_T$-slope due to this dependence are small for all investigated thermal sources, namely for the in-medium hadronic sources (in-medium vector mesons and 
four-pion annihilation) as well as for the partonic source.

In the absence of flow, a dilepton $p_T$ spectrum at given $M$ arises from a superposition of contributions from different $T$, weighted by the volume radiating into the acceptance at this temperature. Since $T$ drops as a result of volume expansion, this means that the low $T$ spectra will typically contribute more to the $p_T$ integrated yield as long as $M$ is sufficiently small, thus adding all contributions the low $p_T$ region will be dominated by late, low $T$ emission whereas at larger $p_T$ the harder slopes of earlier emission from hotter regions will gradually become visible.

The presence of a flow field changes this picture somewhat: Since $T$ and flow are anticorrelated in the model (initially $T$ is large and $\rho_T$ small, at late times the relation is reversed), there is no cold emission source without flow distortion in the model, but hot emission regions are unaffected by flow. For a single temperature, the flow effect leads to a $p_T$-dependent hardening of the spectrum. This blueshift becomes maximal at some $p_T$ (dependent on the values of $M$ and $\rho_T$), the 'knee', and then drops again. The final $p_T$ spectrum arises as the integral over all volume elements with different temperatures and blueshift values, i.e. the knee position is averaged and the $p_T$-dependent sequence of hardening-softening of the slope indicating a 'knee' becomes invisible. The general feature that the soft slopes dominate at low $p_T$ whereas the hardest slope becomes visible at high $p_T$ returns instead (this is, in fact, the major difference of e.m. spectral slopes as compared to hadronic spectral slopes where only contributions from a narrow temperature region need to be considered).
 
The model also imposes a longitudinal flow field, however since it has little relevance for the slope of $p_T$ spectra we do not discuss it further, the interested reader is referred to \cite{Synopsis}.

\section{Comparison with NA60 data}

The NA60 collaboration has measured $p_T$ spectra in three different invariant mass windows: A low mass window with 0.4 GeV $< M < $ 0.6 GeV, the $\rho$-like region with 0.6 GeV $< M <$ 0.9 GeV and the high mass region 1.0 GeV $< M <$ 1.4 GeV. While our model as outlined in \cite{NA60calc} has predicted the general systematics and order of magnitude of the $p_T$ slopes surprisingly well \cite{NA60ptspec} just based on scaling relations from Pb-Pb collisions to the smaller In-In system \cite{NA60calc}, in the following we do the following minor readjustments to improve agreement with data: For the flow field, we use a dependence $\rho \sim \sqrt{r}$ instead of Eq.~(\ref{E-Flow}). In addition, we increase the (unphysical) parameter $v_f$ to $0.6 c$ such that the maximum flow at reached at the medium surface close to $\tau_f$ is about 0.57 $c$. The overall change in the spectral slopes as compared to the results presented in \cite{NA60calc,NA60ptspec} are less than 5\%.

Let us start the discussion with the $\rho$-like mass region. Here, the invariant mass spectra show a clear peak with vacuum width which has been identified with $\rho$ mesons decaying after thermal decoupling of the hot matter in vacuum \cite{NA60calc,NA60hq}. If so, the spectral slope of this contribution should coincide with the spectral slope of a hadronic spectrum at the $\rho$ mass. However, in both analyses \cite{NA60calc,NA60hq} the vacuum contribution does not dominate the yield in the invariant mass window. We may however note that the vacuum contribution is expected to reflect a maximal amount of flow (and hence hardest slope), as it arises from the three-dimensional boundary region of the thermalized four-volume and the flow field is always largest at the edges. Thus, even if this contribution does not dominate when integrated over $p_T$, at sufficiently high $p_T$ it should.

If so, the slope at sufficiently high $p_T$ can be estimated using a phenomenological formula from \cite{OldHad}. There it was shown that the systematics of hadronic spectra at different mass can be described roughly by a fit of the form

\begin{equation}
\label{E-Teff}
\frac{1}{m_T} \frac{dN}{dm_T} = A \exp\left[ -\frac{m_T}{T^*}\right]
\end{equation}

where

\begin{equation}
\label{E-Pocket}
T^* = T + m_{i} \langle v_T \rangle^2
\end{equation}

with $m_i$ the mass of the hadron in question, $T^*$ the apparent spectral slope and $T$ the true temperature of the emission region at freeze-out. Within reasonable limits, this formula reproduces the more microscopic considerations of e.g. \cite{HeinzThermal} quite well. Averaging our flow profile over the fireball breakup time and evaluating this relation, we find $T^* \approx 0.265$ MeV, i.e. a rather hard spectrum. However,  this is in fact not far from the experimentally observed value, given the simple estimate we have made. We take the fact that a peak with vacuum $\rho$ width is observed and that the $p_T$ spectrum in this region approaches the expected value with increasing $p_T$ as a strong indication that the data indeed show the vacuum $\rho$. A decomposition into different sources of the theoretical prediction of the $p_T$-spectrum integrated over 0.6 GeV $< M <$ 0.9 GeV mass region indeed shows that the vacuum-$\rho$ contribution is the dominant source for $p_T \apprge 1$ GeV.
This puts severe limits on the flow field close to $\tau_f$, and by backward extrapolation of the transverse and longitudinal flow fields also on somewhat earlier times.

\begin{figure*}[!htb]
\epsfig{file=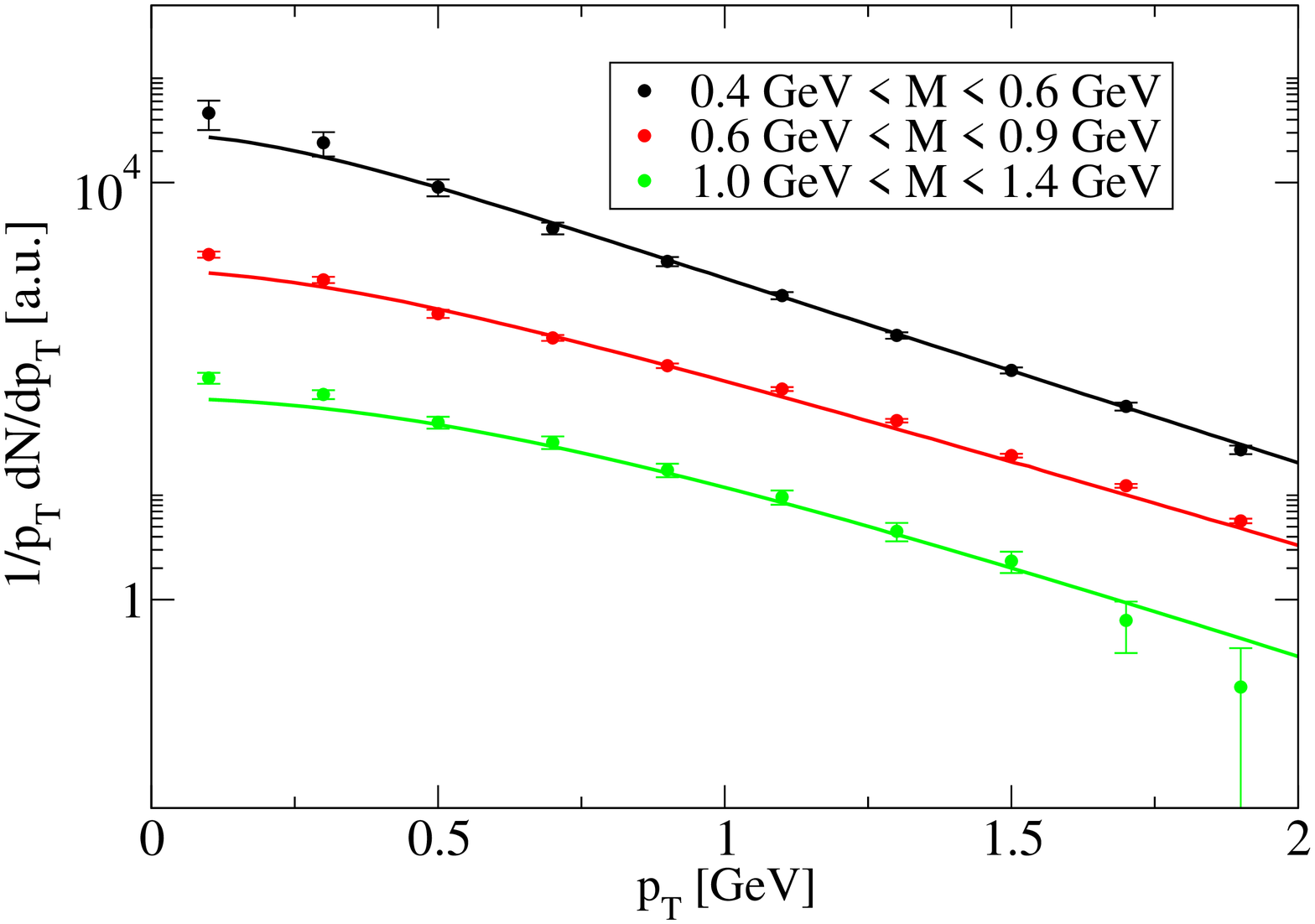, width=7.5cm}\epsfig{file=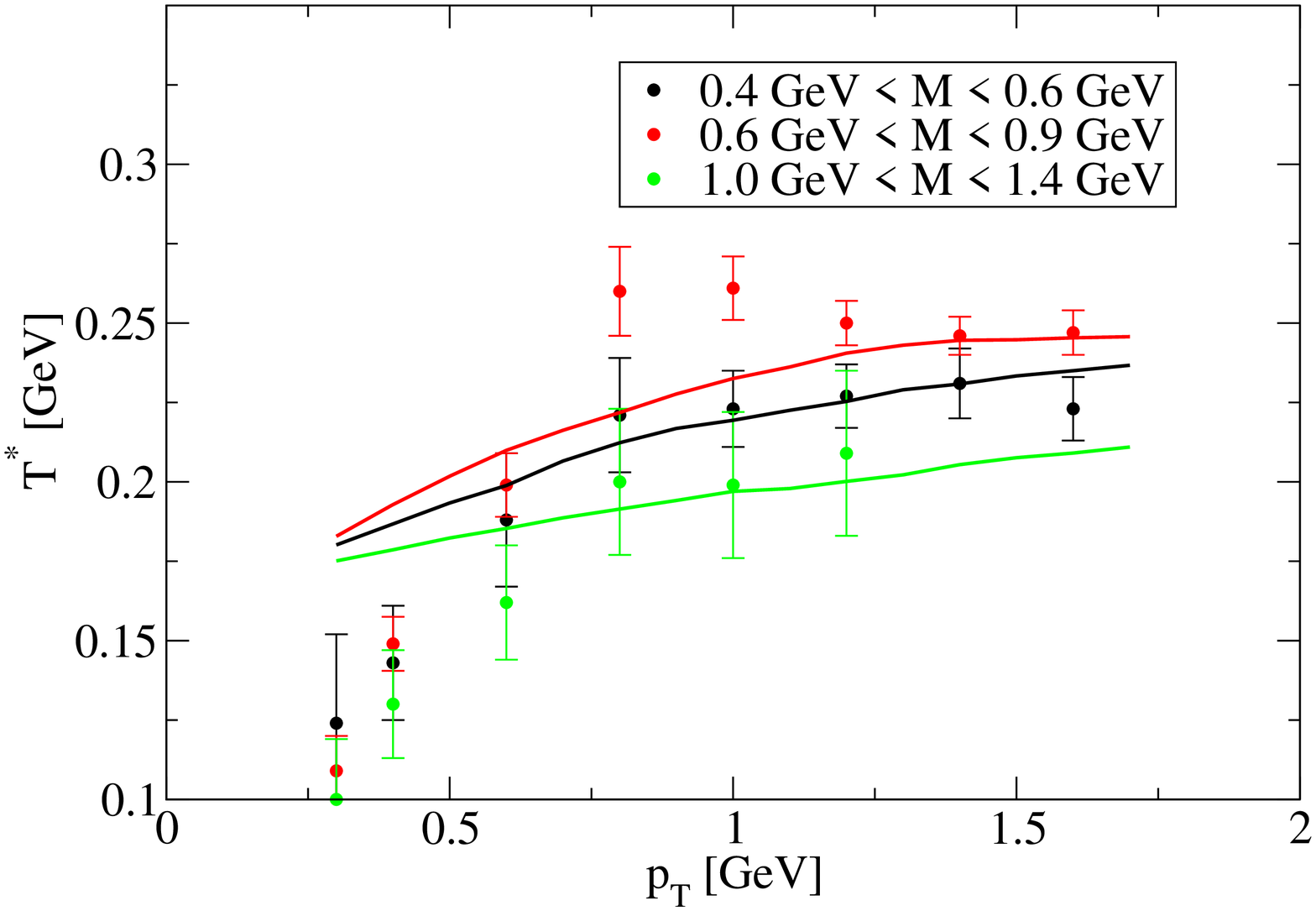, width=7.5cm}

\caption{\label{F-ptcomp}Left panel: $p_T$ spectra in three different mass windows as obtained by NA60 \cite{NA60ptspec} as compared to the model calculation. Right panel: Effective temperature of the spectra determined from an exponential fit for a $p_T$ window of 0.8 GeV (0.6 GeV for the first data point) as a function of the window center.}
\end{figure*}

We present the full calculation of both spectrum and effective temperature $T^*$ as extracted from a local fit of the spectra to the r.h.s. of Eq.~(\ref{E-Teff}) in all three mass windows compared with the data in Fig.~\ref{F-ptcomp}. As can be seen, the calculation with the flow field constrained by the vacuum rho decay describes the spectra in general well except in the low $p_T$ region with $p_T \apprle 0.5$ GeV.

Let us discuss the three mass windows in turn: Both the low mass and the $\rho$ like region receive their dominant contribution in our model from the late hadronic evolution phase, i.e. from regions with temperatures $T < 170$ MeV. This implies that the blueshift of the spectra by flow is large --- it amounts to about 70-80 MeV and is hence even substantially larger than the variation of the actual temperature in the hadronic phase which is only 40 MeV. In particular, it is substantially larger than the uncertainty introduced by the choice of the proper $p_T$ dependence of the spectral function in the hadronic phase --- even a drastic change like neglecting {\em any} $p_T$ dependence of the spectral function does not lead to more than 10 MeV change in slope.

\begin{figure}
\epsfig{file=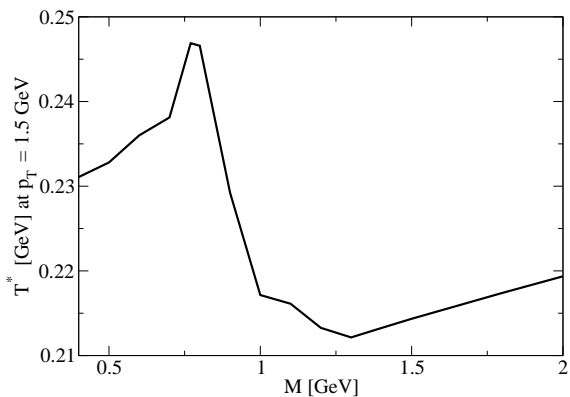, width=7.5cm}
\caption{\label{F-Mdep}Spectral slope $T^*$ at $p_T = 1.5$ GeV calculated as a function of invariant mass $M$.}
\end{figure}

The difference between the low mass and the $\rho$ like-region is essentially caused by the contribution of the vacuum $\rho$ which receives the maximum amount of flow (see above). In contrast (as also apparent from $T^*$) the high mass bin is in our calculation dominated by partonic radiation. This predominantly takes place before the transverse flow field grows large and hence $T^*$ in this momentum window approximately reflects the true temperature of the system. The model results for $T^*$ as a function of $M$ are shown in Fig.~\ref{F-Mdep}. The prominent role of the vacuum rho contribution is clearly apparent from this figure.

\section{The low $p_T$ region}

As apparent from Fig.~\ref{F-ptcomp}, the model fails to reproduce the low $p_T$ region in all three mass bins. Experimentally, values of $T^*$ as low as $\sim 100$ MeV are observed.  Recalling that applying a flow field to a $p_T$ spectrum leads to a blueshift, i.e. higher apparent temperature, we may thus characterize the source radiating into the low $p_T$ region as having a temperature $ T \apprle 100$ MeV and no strong flow. The strict anticorrelation of $T$ and $\rho_T$ in any hydrodynamically motivated model makes it appear very questionable that this could be a thermal source. A similar softening of the low $p_T$ slope is observed e.g. in pion spectra and could there be ascribed to resonance decays \cite{HeinzThermal}, although it is at present unclear what decay process could be responsible for the source dominating the low $p_T$ region in the dimuon spectra. 

Since it can clearly be demonstrated that a thermal description of a flowing medium cannot possible describe the source properly, we will not address the region of $p_T \apprle 0.5$ GeV in the following discussion.

\section{The large $M$ region}

While the data seem to support our idea that the mass region above $1$ GeV is dominated by partonic radiation, the question remains if a hadronic description would not likewise be possible. We will address this question by requiring that the correlation of $\rho_T$ and $T$ must be kept such that the low mass and $\rho$-like region are described properly. Given this constraint, we ask what source characteristics are compatible with the measured slope. 

Note that the fact that $T^*$ is lowest in the high mass region is far from trivial: If the source radiating into high masses were characterized by similar average temperature and flow as the source in the other mass bins, the slope would be expected to be hardest as the blueshift of spectra in a flow field increases with mass (see Eqs.~(\ref{E-TraFo},\ref{E-Pocket})). In fact, if the source had the same average emission temperature, we would expect $T^* = 250 -260$ MeV given our flow field. Thus, it is evident that there must be a transition to a qualitatively different source.

Assuming that the dominant source is hadronic, $T < 170 $ MeV, but not close to $T_F$ (see above) we can next ask how big a window in $T$ with associated changes in the flow field one can allow before the low mass and the $\rho$-like mass bins cannot be described any more by the model. We find that the region from 170 down to 160 MeV is in principle compatible with the data, the region down to 150 MeV is marginally compatible, but the source cannot dominate the high mass yield for $T \apprle 150$ MeV, otherwise the strong flow field (required by the lower mass bins which are dominated by emission from this temperature) pushes $T^*$ above the data. 

Thus, if four-pion annihilation processes were the dominante source like suggested in \cite{RappNA60} they would have to cease contributing rather suddenly around $T \sim 160$ MeV. 

Employing this argument solely based on the $p_T$-slope of the di-muon spectrum, one cannot rule out that the high mass radiation comes dominantly from a hadronic source with temperatures in the range of 150 MeV $\apprle T \apprle$ 170 MeV, although a substantial amount of fine-tuning of the source is required for this assumption.
However, assuming that such a fine-tuning is performed, the measured slope cannot be described by a hadronic source without flow, instead a substantial flow of order $\langle v_T \rangle \apprge 0.2 $ is needed. Such transverse flow indicates a precursor state which cannot be a mixed phase of a 1st order phase transition (as no flow is built up from a mixed phase). Thus, even if the radiation in the high mass window would be of hadronic origin, its slope still points towards a partonic evolution before.

Further constraints can be derived if the fact is taken into account that four-pion annihilation can have strong contributions to the yield from emissions at  temperatures close to $T_f = 130 {\rm MeV} \ll160$ MeV due to the growing fugacity factor $\exp[4 \mu_\pi(T)/T]$  with smaller $T$. In an explicit calculation with such a dimuon source, we employed thermal four-pion annihiliation rates as calculated in the approach \cite{Lichard}. This approach is constrained e.g. by the most recent BaBar data  \cite{BaBar}  on the inverse process in the vacuum, namely lepton annihilation into four-pions.  
The calculation indeed shows that even if fine-tuning of the source is performed to enforce compatibility of $T^*$ with the $p_T$-slope data in the higher mass window, the di-muon yield from these processes is not enough to account for the absolute yield as observed in the mass region for $M>$1.0 GeV assuming these were the dominant source \cite{NA60forth}.

We point out that this investigation only implies that the {\it dominant} source in the $M>$1.0 GeV region cannot be four-pion annihilation processes. It is still possible that these processes
contribute subdominantly to the $M\apprge 1$ GeV dimuon yield in that region of the mass spectrum. 
 A careful analysis of the quantitative strength of such a possible subdominant contribution of four-pion annihilation processes to the spectrum has also to account for the fact that multiparticle annihilation processes can be expected (due to their $\rho_\pi^4$ dependence on the pion density $\rho_\pi$) to decouple earlier than two particle annihilation. Even taking fugacity factors into account, $\rho_\pi$ in the calculation drops by more than a factor two from $T_C$ to $T_F$. While common modelling assumptions treat all processes in kinetic equilibrium down to the same decoupling temperature $T_F$, the experimental findings of momentum equilibration can be accounted for by $2 \leftrightarrow 2$ processes alone, and more microscopic descriptions of the freeze-out process (e.g. hadronic cascades as in \cite{ChihoHydro}) show a sequencial decoupling of different processes. Assuming full equilibrium for multipion annihilation down to $T_F$ thus constitutes an upper limit for the expected contribution.
Further detailed analysis of multipion annihilation is beyond the scope of the present paper, and will be presented elsewhere  \cite{NA60forth}.

\section{Summary}

We have performed an analysis of the $p_T$ spectra by dimuons based on a picture of thermalized matter expanding with a given flow field that has been rather succesful in the analysis of hadronic spectra.  The essentail features of the spectra can be understood from thermal scale and blueshift by flow, thus our results do not crucially depend on details how hadronic or partonic emission is modelled. Requiring that the same flow field and the same $\rho_T -T$ anticorrelation underlies all relevant sources, we have argued that the high mass region 1.0 GeV $<M<1.4$ GeV is most likely dominated by a partonic phase. 

While considerable fine-tuning of the model in principle allows for a hadronic source at $T$ close to the phase boundary $T_C$ in this region, the implication is still that substantial flow is needed to blueshift such a source close to the data which can only arise from a pre-hadronic phase different from a mixed phase. In a thermalized description, this implies  partonic evolution. 
We furthermore have explicilty excluded four-pion annihilation contributions
as such a {\it dominant} hadronic source since its dimuon yield would not be sufficient 
to account solely for the observed yield in the mass spectrum above $M  \apprge 1$ GeV even if the fine-tuning to account for the $p_T$-spectra's slope is performed. 
This does not exclude four-pion annihilation as a subdominant component in the spectrum.
An interpretation of the high mass dilepton radiation in terms of a dominant partonic source seems more natural, as a partonic evolution before hadronic radiation is required by the data in any case.

\begin{acknowledgments}

We would like to thank S.~Damjanovic and H.~Specht for interesting discussions and encouragement to do this investigation. We also thank Peter Lichard
for providing his calculations of four-pion annihilation rates and together with Charles Gale for interesting discussions.
This work was financially supported by the Academy of Finland, Project 206024, and 
by the Natural Sciences and Engineering Research Council of Canada .

\end{acknowledgments}

\end{document}